# On the accurate characterization of business cycles in nonlinear dynamic financial and economic systems.

Dimitri O. Ledenyov and Viktor O. Ledenyov

*Abstract* – The accurate characterization of the business cycles in the nonlinear dynamic financial and economic systems in the time of globalization represents a formidable research problem, which is in the scope of interest by the commercial, investment and central banks. The central banks and other financial institutions make their decisions on the minimum capital requirements, countercyclical capital buffer allocation and capital investments, going from the precise data on the business cycles. The business cycles forecast is a computing intensive mathematical task, which can usually be solved with the application of both the complex algorithms and the parallel computing techniques at the supercomputers. However, in our opinion, the existing computing algorithms need to be improved, because they don't take to the account the fact that the parameters of the business cycles may change during their interactions with the nonlinear dynamic financial and economic systems. We consider the two possible interaction scenarios, when there are 1) the linear interaction, and 2) the non-linear interaction. In our opinion, the main parameters of the business cycle may deviate during the business cycle's nonlinear interaction with the nonlinear dynamic financial and economic systems, because of the origination of the nonlinear effects such as the Four Waves Mixing (FWM), Stimulated Brillouin Scattering (SBS), Stimulated Raman Scattering (SRS), Carrier-Induced Phase Modulation. We emphasis that, in the case of highly nonlinear interaction, the problem of the nonlinear interaction between the business cycles and the nonlinear dynamic financial and economic systems may have the non-stable solutions. Using the econophysical analysis of the nonlinear dynamical financial and economic systems, we propose that, in the frames of new virtuous central bank strategy, there is an increasing necessity to introduce the innovative central bank monetary and financial policies toward the monetary and financial instabilities management in the finances.

PACS numbers: 89.65.Gh, 89.65.-s, 89.75.Fb
Keywords: business cycles, *Juglar* fixed investment cycle, *Kitchin* inventory cycle, *Kondratieff* long wave cycle, *Kuznets* infrastructural investment cycle, econophysics, econometrics, Four Waves Mixing (FWM), Stimulated Brillouin Scattering (SBS), Stimulated Raman Scattering (SRS), Carrier-Induced Phase Modulation nonlinearities, nonlinear dynamic financial and economic systems.

# Introduction

The modern definition of the **business cycle** is given in *Wikipedia (2013)*: "The term *business cycle* (or *economic cycle*) refers to economy-wide fluctuations in production, trade and economic activity in general over several months or years in an economy organized on free-enterprise principles in *Burns, Mitchell (1946)*. These fluctuations occur around a long-term growth trend, and typically involve shifts over time between periods of relatively rapid economic growth (an expansion or boom), and periods of relative stagnation or decline (a contraction or recession). Business cycles are usually measured by considering the growth rate of real gross domestic product. Despite being termed cycles, these fluctuations in economic activity do not follow a mechanical or predictable periodic pattern."

*Juglar (1862)* discovered the **7 –11 years Juglar fixed investment cycle**, which is still in the scope of research interest by many scientists in *Schumpeter (1939)*, *Grinin, Korotayev, Malkov (2010)*, *Korotayev, Tsirel (2010)*, *Ledenyov V O, Ledenyov D O (2012)*, *Ledenyov D O, Ledenyov V O (2013)*. It makes sense to explain that *Schumpeter (1939)* showed that there are the four stages in the *Juglar* cycle: *1)* expansion; *2)* crisis; *3)* recession; *4)* recovery.

*Kitchin (1923)* proposed that there is the **3 – 7 years Kitchin inventory cycle**. This proposition was investigated in *Schumpeter (1939)*, *Korotayev, Tsirel (2010)*, *Ledenyov V O, Ledenyov D O (2012)*, *Ledenyov D O, Ledenyov V O (2013)*.

*Kondratieff (1922, 1925, 1926, 1928, 1935, 1984, 2002)* made a significant contribution to the science of economics. The *Kondratieff's* early research was focused on the big cycles of conjuncture in the World economy in *Kondratieff (1922, 1925, 1926, 1928)*. The discovery of the **45 – 60 years Kondratieff long wave cycle** in *Kondratieff, Stolper (1935)* had a considerable impact on the science of economics. The *Kondratieff's* research achievements are comprehensively analyzed in *Kondratieff (1984, 2002)*. Since that time, the *Kondratieff long wave cycle* has been a subject of intensive research by many scientists in *Schumpeter (1939)*, *Garvy (1943)*, *Silberling (1943)*, *Rostow (1975)*, *Kuczynski (1978, 1982)*, *Forrester (1978, 1981, 1985)*, *Barr (1979)*, *Van Duijn (1979, 1981, 1983)*, *Eklund (1980)*, *Mandel (1980)*, *Van der Zwan (1980)*, *Tinbergen (1981)*, *Van Ewijk (1982)*, *Cleary, Hobbs (1983)*, *Glismann, Rodemer, Wolter (1983)*, *Wallerstein (1984)*, *Bieshaar, Kleinknecht (1984)*, *Zarnowitz (1985)*, *Summers (1986)*, *Freeman (1987)*, *Goldstein (1988)*, *Solomou (1989)*, *Berry (1991)*, *Tylecote (1992)*, *Metz (1992, 1998, 2006)*, *Cooley (1995)*, *Freeman, Louçã (2001)*, *Modelski (2001, 2006)*, *Perez (2002)*, *Rennstich (2002)*, *Rumyantseva (2003)*, *Diebolt, Doliger (2006)*, *Linstone (2006)*, *Thompson (2007)*, *Papenhausen (2008)*, *Taniguchi, Bando, Nakayama (2008)*, *Korotayev, Tsirel*

*(2010), Ikeda, Aoyama, Fujiwara, Iyetomi, Ogimoto, Souma, Yoshikawa (2012), Ledenyov V O, Ledenyov D O (2012), Ledenyov D O, Ledenyov V O (2013).*

*Kuznets (1973)* introduced the **15 – 25 years Kuznets infrastructural investment cycle** in *Kuznets (1973),* based on his research on the cyclical fluctuations of the production and prices in *Kuznets (1930).* The researches on the nature of the Kuznets cycles were conducted by *Abramovitz (1961), Rostow (1975), Solomou (1989); Diebolt, Doliger (2006, 2008), Korotayev, Tsirel (2010), Ledenyov V O, Ledenyov D O (2012), Ledenyov D O, Ledenyov V O (2013).* Most recently, *Korotayev, Tsirel (2010)* conducted the spectral analysis and proposed that there is a tight connection between the *Kondratieff long wave cycle* and the *Kuznets infrastructural investment cycle*, suggesting that the *Kuznets swings* represent a third frequency harmonic of the main frequency oscillation, which is generated by the *Kondratieff long wave cycle*, hence the *Kuznets cycle* is not an independent oscillation in *Korotayev, Tsirel (2010).*

In the macroeconomics, the **multiple origins of business cycles** were proposed: *1) fluctuations in the aggregate demand* in agreement with the *Keynes* theory; *2) fluctuations in the credit* in accordance with the *Minsky* theory; *3) fluctuations in the technological innovations* as explained in the real business cycle theory; *4) fluctuations in the land price* in agreement with the *George* theory in *George (1881, 2009)*; 5) *fluctuations in the politics*. Presently, there are many sophisticated theoretical models, which attempt to explain the origin of *business cycles* in *Schumpeter (1939), Samuelson (1947), Hicks (1950), Inada, Uzawa (1972), Arnord (2002), Sussmuth (2003), Taniguchi, Bando, Nakayama (2008).* It makes sense to explain that, before *2008*, the **great moderation theory** on the decrease of the business cycle's volatility magnitude in the developed countries was very popular among the scientists in *Stock, Watson (2002), Kim Ch-J, Nelson Ch (1999), McConnell, Pérez-Quirós (2000), Bernanke (2004), Summers (2005).*

*Ledenyov V O, Ledenyov D O (2012, 2013)* made the investigations on the **nonlinearities in the economics and finances**. In particular, the authors researched the mixing of various business cycles oscillations, resulting in the origination of the **nonlinear dynamic chaos** in the *nonlinear dynamic financial and economic systems* in the time of globalization in *Ledenyov V O, Ledenyov D O (2012, 2013).*

The main purpose of this research is to understand the **nature of the interaction between the business cycles and the nonlinear dynamic financial and economic systems** with the purpose to precisely characterize the *amplitude*, *phase* and *frequency* of the *business cycle*, making it possible to perform the accurate economic forecasts by the central banks and other financial institutions. Authors use the *econophysics* approach to make the advanced research in the finances and economics, utilizing the knowledge base on the interaction between the electromagnetic waves (light) and the nonlinear media, which can have place in the microwave resonators, optical resonators, fibers and crystals in *Ledenyov D O, Ledenyov V O (2012, 2013)*, *Dutton (1998)*.

**Business cycles detection, filtering and measurement**

"The time dependence of real **Gross Domestic Product (GDP)** usually consists of the fluctuations under the long term growth period," in *Taniguchi, Bando, Nakayama (2008)*. The fluctuations of the *GDP* have the cyclical nature. Making the research on the *business cycle* in the economics, *Taniguchi, Bando, Nakayama (2008)* state: "The observed data show that there exist the *business cycles* in the economies of almost all modern countries in the world. It is a quite general feature of economic system, and has long been one of the most interesting questions how to explain such business cycle." *Taniguchi, Bando, Nakayama (2008)* introduced the following expression: $\Delta G(i) = \Delta G(i) - \Delta G(i-1)$, with the purpose to accurately characterize the *business cycle*. *Taniguchi, Bando, Nakayama (2008)* continue: "Such business cycle may be classified into several types according to characteristics, especially its period: the *Kitchin inventory cycle*, the *Juglar fixed investment cycle*, the *Kuznets infrastructural investment cycle* and the *Kondratieff wave*, which have the period *3 – 5, 7 –11, 15 – 25*, and *45 – 60* years respectively."

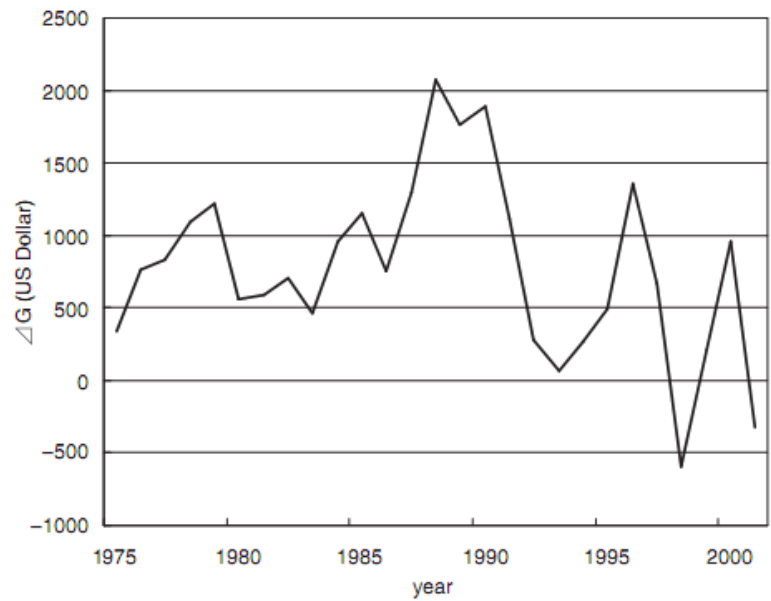

*Fig. 1. Dependence of ΔG(i), which is calculated from GDP per capita (constant 1995 US dollar), shows presence of business cycle with periods of depression and prosperity in Japan (after Taniguchi, Bando, Nakayama (2008).*

Let us consider the *business cycle's* characteristics comprehensively. In the simplified case, the *business cycle* can be represented as a periodic continuous wave oscillation, which is characterized by the *amplitude*, *frequency*, and *phase* parameters. Going from the characteristics of *business cycles* such as their *amplitudes*, *frequencies* and *phases*, the five main types of *business cycles* are distinguished in *Juglar (1862)*, *Kitchin (1923)*, *Kondratieff, Stolper (1935)*, *Kuznets (1973)*, *Taniguchi, Bando, Nakayama (2008)*, *Ikeda, Aoyama, Fujiwara, Iyetomi, Ogimoto, Souma, Yoshikawa (2012)*:

1. *The 3 – 7 years **Kitchin inventory cycle** in Kitchin (1923);*
2. *The 7 –11 years **Juglar fixed investment cycle** in Juglar (1862);*
3. *The 15 – 25 years **Kuznets infrastructural investment cycle** in Kuznets (1973);*
4. *The 45 – 60 years **Kondratieff long wave cycle** in Kondratieff, Stolper (1935); and*
5. *The 70+ **Grand super-cycle**.*

The *business cycle* is considered as a macroeconomic phenomenon in *Bernanke (1979)*. In every economy, there are many economic variables, which can fluctuate over the time. *Bernanke (1979)*, who made the original researches on the long-term commitments, dynamic optimization, and the business cycle, writes: "Economic theorists are usually willing to assume the existence of a great deal of flexibility in the economy. Factors are mobile, prices shift readily, techniques of production are changed, the capital stock is as easily decreases as increased." *Bernanke (1979)* continues: "Analysis of the *business cycle* – a short to medium term phenomenon – needs to recognize the difficulty the economy may have in adjusting to new events." Considering his investment decision model, *Bernanke (1979)* explains that the volatility in the investment demand has the cyclical fluctuations nature. Moreover, *Bernanke (1979)* argues that the investment decisions depend on the *business cycle* strongly.

*Korotayev, Tsirel (2010)* completed their advanced research on the spectral analysis of world *GDP* dynamics with the particular interest in the *Kondratieff*, *Kuznets*, *Juglar* and *Kitchin* cycles spectroscopy in the conditions of global economic development. The main aim of spectral analysis was to detect the presence cyclic fluctuations in the world *GDP* dynamics The completed spectral analysis evidently confirmed that there are the *Juglar, Kitchin, Kondratieff cycles* in the world *GDP* dynamics, suggesting that the *Kuznets cycle* has to be regarded as the third harmonic of the *Kondratieff* wave rather than as a separate independent cycle.

In Fig. 2, the dynamics of World *GDP* annual growth rates (%) in *1871–2007* is shown in *Korotayev, Tsirel (2010)*, *Maddison (1995, 2001, 2003, 2009)*, *World Bank (2009a)*.

In Fig. 3, the *Kondratieff* waves and U.S. wholesale prices are pictured in *Dickson (1983)*, *Korotayev, Tsirel (2010)*.

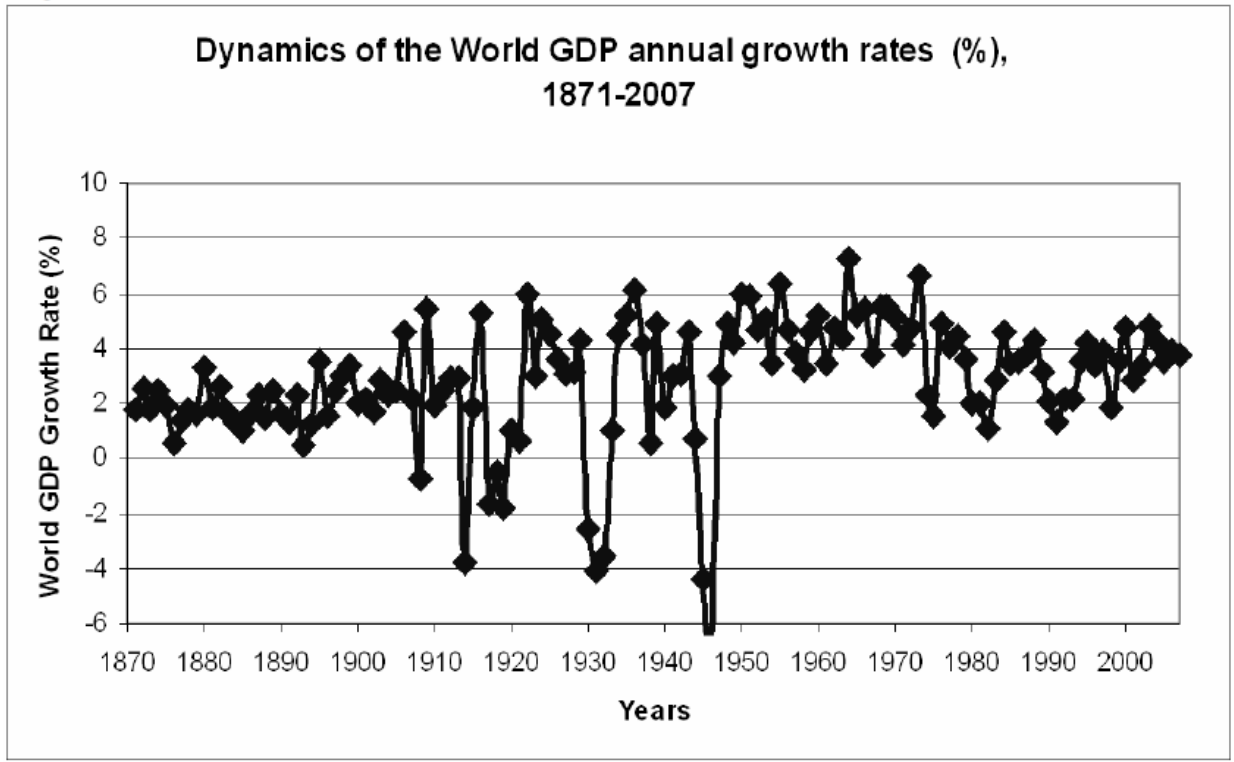

***Fig. 2.*** *Dynamics of World GDP annual growth rates (%) in 1871–2007 (after Korotayev, Tsirel (2010), Maddison (1995, 2001, 2003, 2009), World Bank (2009a)).*

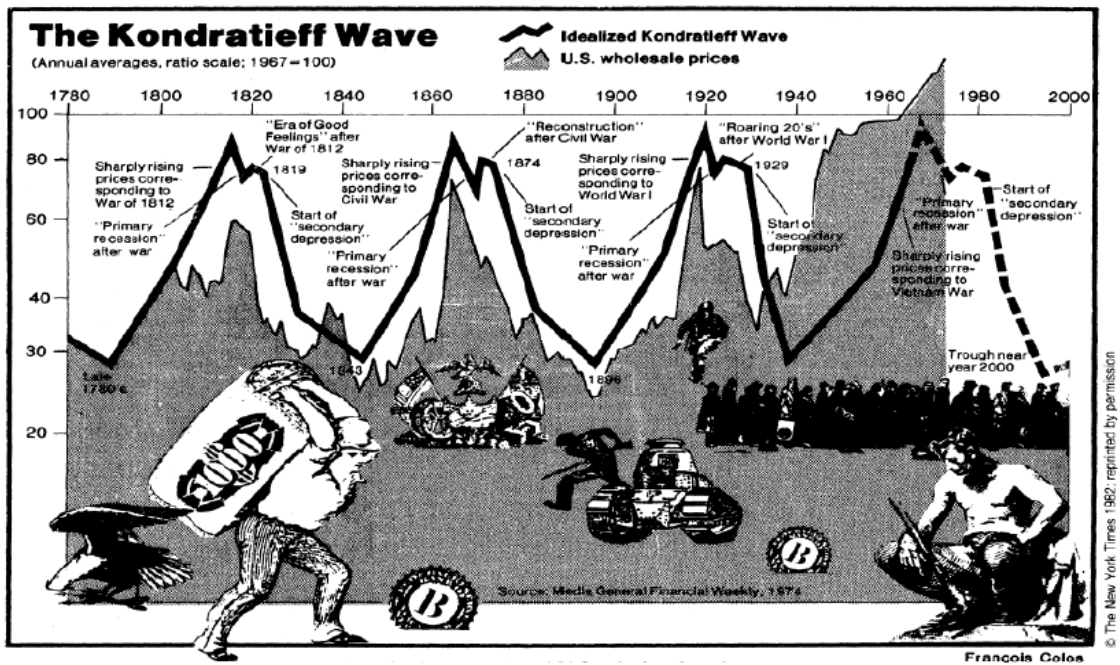

***Fig. 3.*** *Kondratieff waves and U.S. wholesale prices (after Dickson (1983), Korotayev, Tsirel (2010)).*

In Tabs. 1 and 2, *Korotayev, Tsirel (2010)* show the long waves and their phases, which were identified by *Kondratieff* and the "*Post-Kondratieff*" long waves and their phases.

In Tab. 3 and Fig. 4, *Korotayev, Tsirel (2010)* provide the information on the average annual World *GDP* growth rates (%) during phases *A* and *B* of *Kondratieff* waves in *1871–2007*.

| Long wave number | Long wave phase | Dates of the beginning | Dates of the end |
|---|---|---|---|
| One | A: upswing | "The end of the 1780s or beginning of the 1790s" | 1810–1817 |
| | B: downswing | 1810–1817 | 1844–1851 |
| Two | A: upswing | 1844–1851 | 1870–1875 |
| | B: downswing | 1870–1875 | 1890–1896 |
| Three | A: upswing | 1890–1896 | 1914–1920 |
| | B: downswing | 1914–1920 | |

*Tab. 1.* Long Waves and Their Phases Identified by Kondratieff (after Korotayev, Tsirel (2010)).

| Long wave number | Long wave phase | Dates of the beginning | Dates of the end |
|---|---|---|---|
| Three | A: upswing | 1890–1896 | 1914–1920 |
| | B: downswing | From 1914 to 1928/29 | 1939–1950 |
| Four | A: upswing | 1939–1950 | 1968–1974 |
| | B: downswing | 1968–1974 | 1984–1991 |
| Five | A: upswing | 1984–1991 | 2008–2010? |
| | B: downswing | 2008–2010? | ? |

*Tab. 2.* "Post-Kondratieff" Long Waves and Their Phases (after Korotayev, Tsirel (2010)).

| Kondratieff wave number | Phase | Years Version 1 | Years Version 2 | Average annual World GDP growth rates (%) during respective phase Version 1 | Version 2 |
|---|---|---|---|---|---|
| II | End of Phase A | 1871–1875 | 1871–1875 | 2.09 | 2.09 |
| II | B | 1876–1894 | 1876–1894 | 1.68 | 1.68 |
| III | A | 1895–1913 | 1895–1929 | 2.57 | 2.34 |
| III | B | 1914–1946 | 1930–1946 | 1.50 | 0.98 |
| IV | A | 1947–1973 | 1947–1973 | 4.84 | 4.84 |
| IV | B | 1974–1991 | 1974–1983 | 3.05 | 2.88 |
| V | A | 1992–2007 | 1984–2007 | 3.49 | 3.42 |

*Tab. 3.* Average annual World GDP growth rates (%) during phases A and B of Kondratieff waves in 1871–2007 (after Korotayev, Tsirel (2010)).

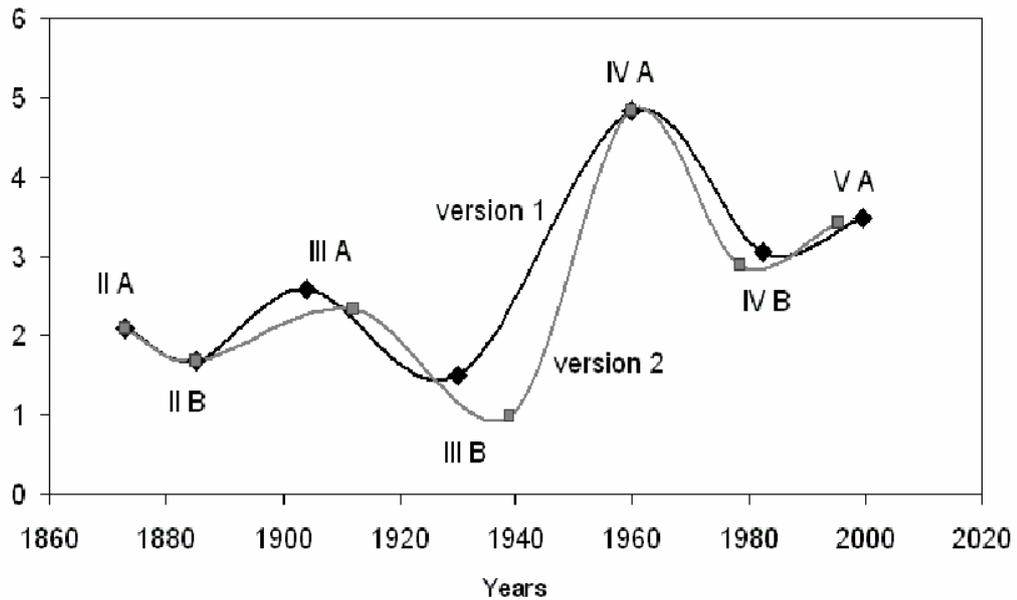

*Fig. 4. Average annual World GDP growth rates (%) during phases A and B of Kondratieff waves in 1871–2007 (after Korotayev, Tsirel (2010)).*

*Baxter, King (1999)* made the research titled: "Measuring business cycles: approximate band-pass filters for economic time series," considering the application of various signal filtering techniques to detect the *business cycles* in the economics. We would like to demonstrate the modern mathematical technique to analyse the oscillations of *GNP*, which originate the *business cycles*. Using the research data in *Hodrick, Prescott (1997), U.S. Federal Reserve Economic Data FRED (2012)*, it is possible to plot the cyclical component of *GNP* in *Matlab R2012 (2012)*. Let us provide an example, which shows how to use the *Hodrick-Prescott filter* to decompose a time series in *Matlab R2012 (2012)*. The *Hodrick-Prescott filter* transforms a time series into the cyclical components in *Matlab R2012 (2012)*

$$y_t = g_t + c_t$$

where $y_t$ is a time series, $g_t$ is the growth component of $y_t$, and $c_t$ is the cyclical component of $y_t$ for $t = 1, \ldots, T$.

The objective function for the *Hodrick-Prescott filter* has the form in *Matlab R2012 (2012)*

$$\sum_{t=1}^{T} c_t^2 + \lambda \sum_{t=2}^{T-1} \left( \left( g_{t+1} - g_t \right) - \left( g_t - g_{t-1} \right) \right)^2$$

with the smoothing parameter *lambda*. The programming problem is to minimize the objective over all the $g_1, \ldots, g_T$ in *Matlab R2012 (2012)*. The conceptual basis for this programming problem is that the first sum minimizes the difference between the data and its growth component (which is the cyclical component) and the second sum minimizes the second-order difference of the growth component, which is analogous to minimization of the second derivative of the growth component in *Matlab R2012 (2012)*.

In Fig. 5, the plot of cyclical *GNP*, created with the *Hodrick-Prescott* filter to analyze *GNP* cyclicality is shown *in Hodrick, Prescott (1997), U.S. Federal Reserve Economic Data FRED (2012), Matlab R2012 (2012))*.

As explained in *Matlab R2012 (2012)*, the blue line is a cyclical component with the smoothing parameter *1600* and the red line is a difference with the respect to the de-trended cyclical component. The difference is smooth enough to suggest that the choice of smoothing parameter is appropriate in *Matlab R2012 (2012)*. There are some differences in the graphs due to the differences in the sources for the pair of adjustments since the *GNP* data in Fig. 5 and in the research paper in *Hodrick, Prescott (1997)* are both adjusted for the seasonal variations with the conversion from the nominal to the real values. The research statistical data comes from the *U.S. Federal Reserve Economic Data FRED (2012)*.

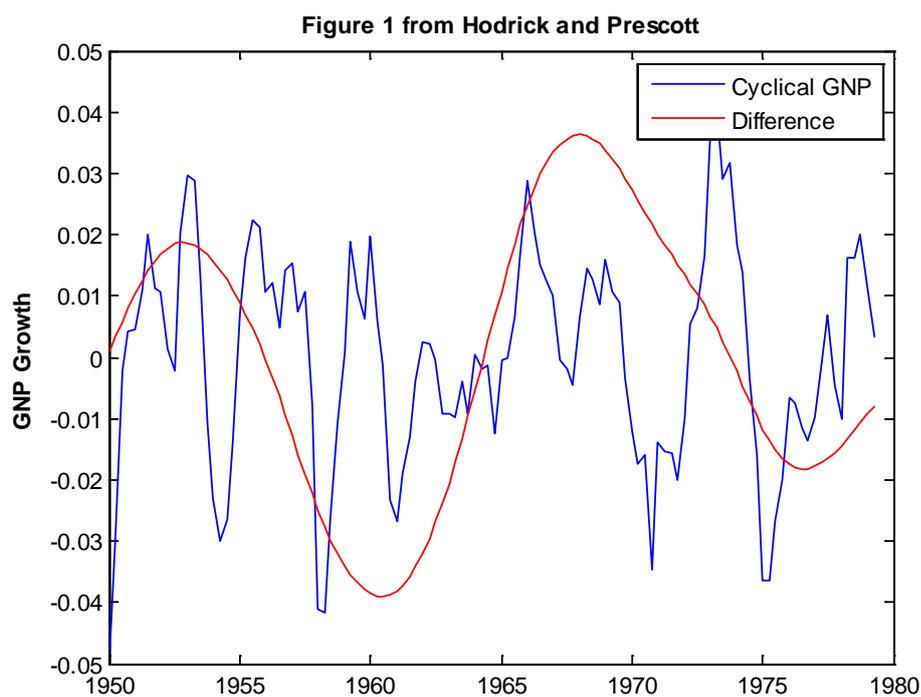

***Fig. 5.*** *Plot of cyclical GNP, created with Hodrick-Prescott filter to analyze GNP cyclicality (after Hodrick, Prescott (1997), U.S. Federal Reserve Economic Data FRED (2012), Matlab R2012 (2012))*.

Most recently, *Ikeda, Aoyama, Fujiwara, Iyetomi, Ogimoto, Souma, Yoshikawa (2012)* researched the coupled oscillator model of the *business cycle* as shown in Figs. 6, 7, 8.

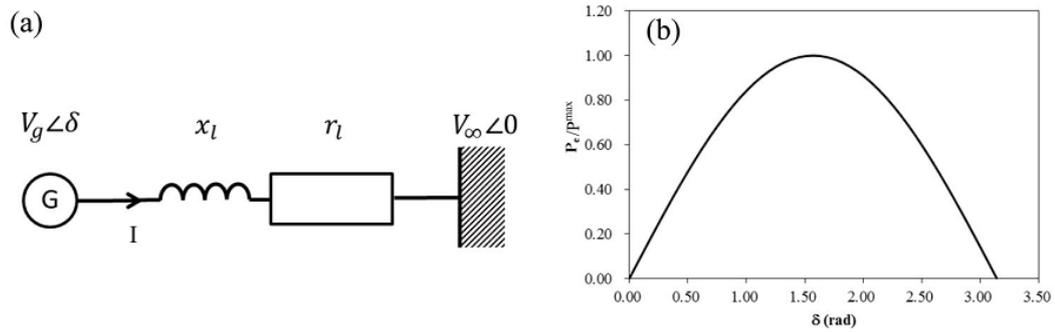

***Fig. 6.*** *Single machine connected to infinite bus and synchronizing force (after Ikeda, Aoyama, Fujiwara, Iyetomi, Ogimoto, Souma, Yoshikawa (2012)).*

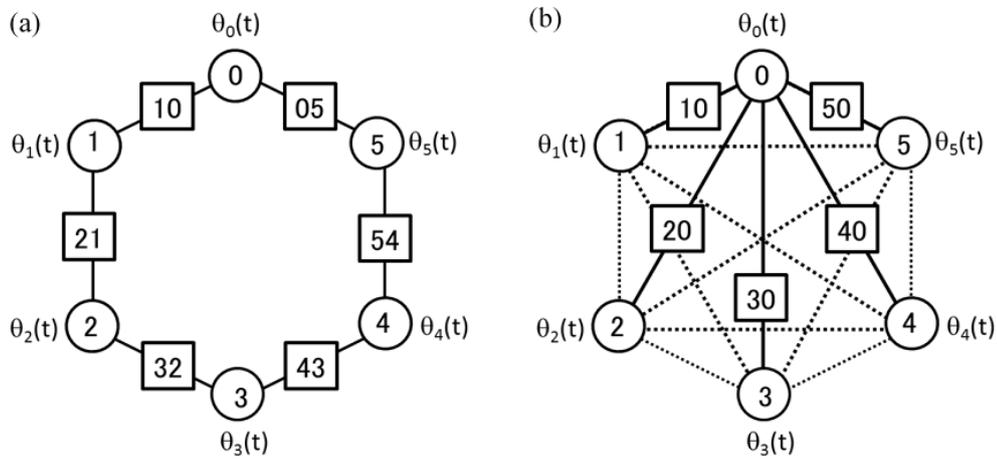

***Fig. 7.*** *System consisting of six oscillators: (a) the Nearest Neighbor (NN) graph; (b) the Complete (C) graph; oscillators and goods markets are indicated by circles and rectangles, respectively (after Ikeda, Aoyama, Fujiwara, Iyetomi, Ogimoto, Souma, Yoshikawa (2012)).*

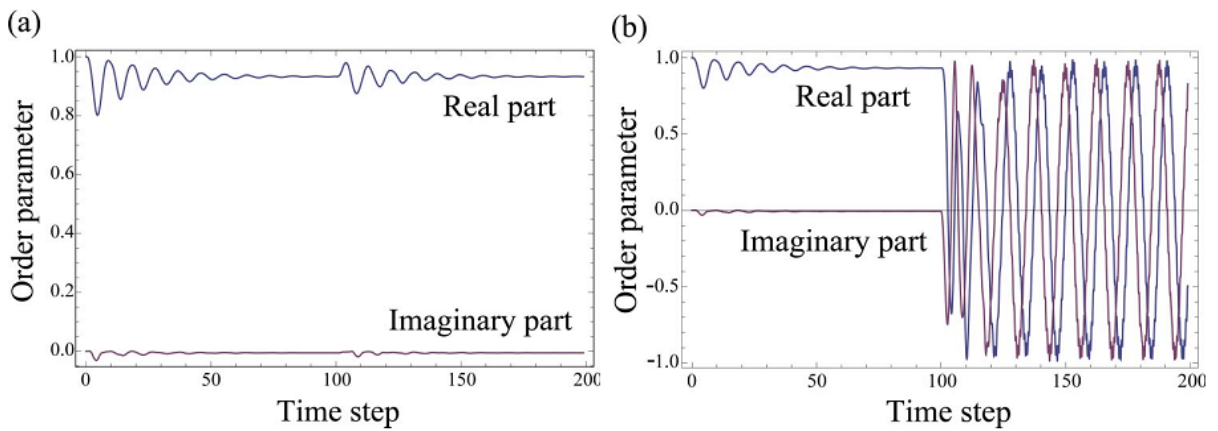

***Fig. 8.*** *Price elasticity and stability of synchronization in fluctuating goods market (after Ikeda, Aoyama, Fujiwara, Iyetomi, Ogimoto, Souma, Yoshikawa (2012)).*

# Interaction of business cycles with nonlinear dynamic financial and economic systems

"The *business cycle* is observed in most of industrialized economies. Economists have studied this phenomenon by means of mathematical models, including various kinds of linear non-linear, and coupled oscillator model," as it is explained in *Ikeda, Aoyama, Fujiwara, Iyetomi, Ogimoto, Souma, Yoshikawa (2012)*. In addition, *Ikeda, Aoyama, Fujiwara, Iyetomi, Ogimoto, Souma, Yoshikawa (2012)* write: "The *business cycle* is an example of synchronization, and this has been studied in *nonlinear physics*."

We will use the *econophysics* approach to research the nature of interaction between the *business cycles* and the *nonlinear dynamic financial and economic systems* with the aim to precisely characterize the *amplitude*, *phase* and *frequency* of the *business cycle* in the economics and finances. Therefore, let us describe the **nonlinear effects** in the field of physics, and then search for the possible similarities in the economics and finances.

It is a well known fact in the optical physics that the different wavelengths of the light do not interact with one another, when the light propagates in the vacuum in *Dutton (1998)*. However, the different wavelengths of the light can interact with one another, when the light propagates in the optical fibers or crystals in *Dutton (1998)*. This interaction can originate the changes of characteristics of the light wave itself, including the amplitude, wavelength, and phase in *Dutton (1998)*. Moreover, the interaction of the light with the material can have the linear or non-linear nature in *Dutton (1998)*. At the increased levels of light power, the following **nonlinear effects**, which have the exponential dependences, can originate in *Dutton (1998)*:

1. The **Four Waves Mixing (FWM) effect**, when the two different light wavelengths can mix together to generate the new signals at the wavelengths, which are spaced at the same intervals as the mixing signals. The light signal with the frequency of $\omega_1$ mixes with the light signal with the frequency $\omega_2$ to generate the two new signals at the frequency of $2\omega_1-\omega_2$ and at the frequency of $2\omega_2-\omega_1$.
2. The **Stimulated Brillouin Scattering (SBS) effect**, when the forward propagating light scatters backward, because of the acoustic vibrations in the optical fiber. The *BCS* effect is caused by the light being reflected by the diffraction grating created by the regular pattern of the refraction index changes in the optical fiber or crystal. The reflected light is reflected backward from a moving grating, hence its frequency is shifted by the *Doppler* effect.

3. The ***Stimulated Raman Scattering (SRS) effect***, when the forward propagating light scatters backward, because of the molecular vibrations in the optical fiber. The power of light signal with the shorter wavelength can transfer to the light signal with the longer wavelength in the optical fiber.

4. The ***Carrier-Induced Phase Modulation effect***, which includes both the *Self Phase Modulation (SPM)* and *Cross Phase Modulation (XPM)*, when the presence of light in a fiber causes a tiny change in the refractive index of the fiber, because of the action by the light on the atoms and molecules of the glass (plastic)- the *Kerr* effect, resulting in the *SPM* and *XPM*. There are the *linear Kerr effect* and *nonlinear Kerr effect*, depending on the intensity of the applied light in the fiber. The *SPM* causes the change of phase of the light waves in the pulse, because of the *RI* difference at the leading edge, trailing edge, middle of the light waves pulse in the fiber. Therefore, the frequency spectrum of the pulse is broadened. In the case of *XPM*, there are multiple signals at different wavelengths in the same fiber and the Kerr effect caused by one signal can result in the phase modulation of the other signal(s). There is no power transfer between the optical signals at the *XPM*, but there may be the asymmetric spectral broadening and distortion of the pulse shape.

Thus, we can see that the interaction between the light waves and the nonlinear medium may result in an origination of the four well known nonlinear effects such as ***Stimulated Brillouin Scattering (SBS), Stimulated Brillouin Scattering (SBS), Stimulated Raman Scattering (SRS)*** and ***Carrier-Induced Phase Modulation***.

In our opinion, the interaction between the *business cycles* and the *nonlinear dynamic financial and economic systems* in the economics and finances is not researched yet. Therefore, we propose to consider the *nonlinear dynamic financial and economic systems* as a nonlinear medium. The interaction between the *business cycles* (the *business cycle* has its amplitude, frequency and phase) and the *nonlinear dynamic financial and economic systems* can be described by the two possible types of interactions:

1. The ***linear interaction***: The output sinusoid signal (the *business cycle*) has the same frequency, but may undergo the amplitude and phase changes, without the additional signals creation;

2. The ***non-linear interaction***: The output signal (the *business cycle*) has the shifted frequency and may undergo the amplitude and phase changes, with the additional signals creation such as the harmonics or the inter-modulation products.

*We make the theoretical proposition that the following four nonlinear effects may originate as a result of the nonlinear interaction between the business cycles and the nonlinear dynamic financial and economic system*:

1. ***Four Waves Mixing (FWM) effect,*** when the two different business cycles wavelengths can mix together to generate the new signals at the wavelengths, which are spaced at the same intervals as the mixing signals in the *nonlinear dynamic financial and economic system*s. The business cycle's oscillation with the frequency of $\omega_1$ mixes with the business cycle's oscillation with the frequency $\omega_2$ to generate the two new business cycle's oscillations at the frequency of $2\omega_1-\omega_2$ and at the frequency of $2\omega_2-\omega_1$ in the *nonlinear dynamic financial and economic systems*.

2. ***Stimulated Brillouin Scattering (SBS) effect,*** when the *forward propagating business cycle's oscillatory signal of certain wavelength* scatters backward, because of the acoustic-like vibrations in the *nonlinear dynamic financial and economic systems*. The *BCS* effect may be caused by the *forward propagating business cycle's oscillatory signal of certain wavelength* being reflected by the diffraction grating, created by the regular pattern of the refraction index changes in the *nonlinear dynamic financial and economic systems*. It is necessary to note that the *reflected business cycle's oscillatory signal of certain wavelength* is reflected backward from a moving diffraction grating, hence its frequency will be shifted by the *Doppler* effect.

3. ***Stimulated Raman Scattering (SRS) effect***, when the *forward propagating business cycle's oscillatory signal of certain wavelength* scatters backward, because of the molecular-like vibrations in the *nonlinear dynamic financial and economic systems*. The power of the *business cycle's oscillatory signal with the shorter wavelength* can transfer to the *business cycle's oscillatory signal with the longer wavelength* in the *nonlinear dynamic financial and economic systems*.

4. ***Carrier-Induced Phase Modulation effect***, which includes both the *Self Phase Modulation (SPM)* and *Cross Phase Modulation (XPM)*, when the presence of the *forward propagating business cycle's oscillatory signal of certain wavelength* in the *nonlinear dynamic financial and economic systems* causes a tiny change in the refractive index of the *nonlinear dynamic financial and economic systems*, because of the action by the business cycle on the economic agents in the *nonlinear dynamic financial and economic systems* - the *Kerr* effect in the *nonlinear dynamic financial and economic system*s, resulting in the origination of the *SPM* and *XPM*. There may be the *linear Kerr effect* and *nonlinear Kerr effect*, depending on the intensity of the applied light in the in the *nonlinear dynamic financial and economic system*s. The *SPM* causes the change of phase of the *forward propagating business cycle's waves* in the pulse, because of the *RI* difference at the leading edge, trailing edge, middle of the light waves pulse in the *nonlinear dynamic financial and economic system*s. Therefore, the frequency spectrum of the pulse is broadened. In the case of *XPM*, there are multiple signals at different wavelengths in the same fiber and the *Kerr* effect, caused by one signal can result in the phase modulation of the other signal(s). There is no power transfer between the *forward propagating business cycle's oscillatory signals* at the *XPM*, but there may be the asymmetric spectral broadening and distortion of the pulse shape.

## Central bank monetary and financial policies improvement due to accurate characterizations of business cycles

The **central bank** is the principal monetary authority of a nation, which performs several key functions, including conducting monetary policy to stabilize the economy and level of prices in *Fox, Alvarez, Braunstein, Emerson, Johnson, Johnson, Malphrus, Reinhart, Roseman, Spillenkothen, Stockton (2005)*.

The **Federal Reserve System** is the central bank of the United States in *Fox, Alvarez, Braunstein, Emerson, Johnson, Johnson, Malphrus, Reinhart, Roseman, Spillenkothen, Stockton (2005)*. It was founded by *Congress* in *1913* to provide the nation with a safer, more flexible, and more stable monetary and financial system in *Fox, Alvarez, Braunstein, Emerson, Johnson, Johnson, Malphrus, Reinhart, Roseman, Spillenkothen, Stockton (2005)*. The *Federal Reserve System's* duties fall into four general areas in *Fox, Alvarez, Braunstein, Emerson, Johnson, Johnson, Malphrus, Reinhart, Roseman, Spillenkothen, Stockton (2005)*:

1. Conducting the nation's monetary policy by influencing the monetary and credit conditions in the economy in pursuit of maximum employment, stable prices, and moderate long-term interest rates.
2. Supervising and regulating banking institutions to ensure the safety and soundness of the nation's banking and financial system and to protect the credit rights of consumers.
3. Maintaining the stability of the financial system and containing systemic risk that may arise in financial markets.

4. Providing financial services to depository institutions, the U.S. government, and foreign official institutions, including playing a major role in operating the nation's payments system.

In the time of globalization, the accurate characterization of the *business cycles* in the *nonlinear dynamic financial and economic systems* represents a formidable challenge for the commercial, investment and *central banks*. The *central banks* develop the **monetary and financial policies**, including the *minimum capital requirements* and *countercyclical capital buffer requirements*, with the use of the data on the *business cycles* in *Bernanke, Blinder (1992), Bernanke, Gertler (1995), Bernanke (1995, 2004, 2007, 2009, 2010, 2013), Bernanke, Blanchard, Summers, Weber (2013), Swiss National Bank (2013), Central Banking Newsdesk (2013)*. Moreover, the inability to accurately characterize the *business cycles* may result in the execution of very wrong central bank polices and the subsequent strong necessity for the introduction of the crisis management policies, including the classic prescriptions of *liquidity provision*, *liability guarantees*, *asset re-evaluation* and *disposition*, and *re-capitalization* in *Bernanke (2013)*. In addition, the commercial and investment banks make the capital investment and budgeting decisions, going from the obtained precise data on the *business cycles* in *Bernanke, Blanchard, Summers, Weber (2013)*.

*Drehmann, Borio, Tsatsaronis (2011)* mention: "Financial boom-and-bust cycles are costly for the banks involved and for the economy at large." *Drehmann, Borio, Tsatsaronis (2011)* continue to explain: "… the experience added impetus to policymakers' and academic economists' efforts to better understand the mechanisms that drive financial system pro-cyclicality and to devise policy tools that can mitigate it." *Drehmann, Borio, Tsatsaronis (2011)* add that the time-varying regulatory capital buffers for banks can provide a reliable guide for regulatory capital requirements to dampen banks' pro-cyclical behavior, restraining risk taking during booms and cushioning financial distress during busts. It is necessary to explain that *Drehmann, Borio, Tsatsaronis (2011)* distinguish the three objectives of countercyclical policies:

1. To smooth the *business cycle* through the influence of capital requirements on banks, i.e., to use them as a demand management tool. Setting such an objective is commensurate to calibrating time-varying prudential capital requirements to achieve a *macroeconomic goal*.
2. To smooth the *financial (credit) cycle*. We think of this approach as using capital requirements as a means of achieving a *broad macroprudential goal*.

3. To protect the banks from the build-up of system-wide vulnerabilities. We call this a *narrow macroprudential goal*, designed to strengthen systemic resilience without taking explicit account of its influence on the financial and business cycles.

We would like to state that the *business cycles forecast* is a computing intensive mathematical task, which can usually be solved with the application of the parallel computing techniques at the supercomputers. In our opinion, the existing computational algorithms need to be significantly improved, because they don't take to the account the fact that the main parameters of the *business cycle* may deviate during its nonlinear interactions with the nonlinear medium such as the *nonlinear dynamic financial and economic systems*, because of the **Four Waves Mixing (FWM), Stimulated Brillouin Scattering (SBS), Stimulated Raman Scattering (SRS), Carrier-Induced Phase Modulation effects**. We think that the nonlinear interaction between the *business cycles* and the *nonlinear dynamic financial and economic systems*, results in both the changes of the business cycles *amplitude*, *phase* and *frequency* and the appearance of harmonics and inter-modulation products (see the book on the nonlinearities in microwave superconductivity in *D O Ledenyov, V O Ledenyov (2012)*).

We considered the two possible scenarios of the interaction between the *business cycle* and the *nonlinear dynamic financial and economic systems*: *1)* the *linear interaction*, and *2)* the *non-linear interaction*.

We argue that the described changes in the *business cycle* characteristics, caused by the nonlinear interaction between the *business cycle* and the *nonlinear dynamic financial and economic systems*, have to be taken to the consideration during the process of accurate characterization of the *business cycle* by the central banks.

We would like to comment that, in the case of highly nonlinear interaction, the problem of the nonlinear interaction between the *business cycles* and the *nonlinear dynamic financial and economic systems* may have the non-stable solutions in *D. O. Ledenyov, V O Ledenyov (2013)*. The estimation of *stability* of the financial and monetary systems is not an easy task in *Swiss National Bank (2012)*, because it is increasingly difficult to estimate the *stability* of the *nonlinear dynamic financial and economic systems* in view of the fact that there is no a general approach on the *stability* assessment in the case of the *nonlinear systems*, comparing to the linear systems as explained in *Adhikari (2013)*. Of course, it is well known that: "There is a perception that the *stability* is a basic condition for the practically useful systems, because the unstable systems just can be thought as useless" in *Adhikari (2013)*. However, in our opinion, the inherently unstable systems have one very important advantage, namely these unstable systems can instantly react on the constantly changing external conditions. For example, the modern jets and some other aerospace systems are inherently unstable and they use the near real time *fly-by-wire* concept to react promptly on a sequence of fast external changes in the aerospace. Therefore, going from the *econophysical analysis* of the *nonlinear dynamical financial and economic systems*, we propose that, in the frames of new central bank strategy, there is an increasing necessity to introduce the innovative central bank monetary and financial policies toward the monetary and financial instabilities management, based on the data about the accurately characterized business cycles, in the finances in *Ledenyov D O, Ledenyov V O (2012)*.

**Conclusion**

We would like to conclude by saying that the accurate characterization of the *business cycles* in the *nonlinear dynamic financial and economic systems* in the time of globalization represents a formidable research problem, which is in the scope of interest by the commercial, investment and central banks. The central banks and other financial institutions make their decisions on the minimum capital requirements, countercyclical capital buffer allocation and capital investments, going from the precise data on the business cycles. The *business cycles forecast* is a computing intensive mathematical task, which can usually be solved with the application of both the complex algorithms and the parallel computing techniques at the supercomputers. However, in our opinion, the existing computing algorithms need to be improved, because they don't take to the account the fact that the parameters of the business cycles may change during their interactions with the nonlinear dynamic financial and economic systems. In the presented research, we consider the two possible interaction scenarios, when there are 1) the linear interaction, and 2) the non-linear interaction. In our opinion, the main parameters of the business cycle may deviate during the business cycle's nonlinear interaction with the nonlinear dynamic financial and economic systems, because of the origination of the nonlinear effects such as the *Four Waves Mixing (FWM), Stimulated Brillouin Scattering (SBS), Stimulated Raman Scattering (SRS), Carrier-Induced Phase Modulation*. These changes have to be taken to the account during the precise characterization of the *business cycles*. We emphasis that, in the case of highly nonlinear interaction, the problem of the nonlinear interaction between the business cycles and the nonlinear dynamic financial and economic systems may have the non-stable solutions. Using the *econophysical analysis* of the *nonlinear dynamical financial and economic systems*, we propose that, in the frames of new virtuous central bank strategy, there is an increasing necessity to introduce the innovative central bank monetary and financial policies toward the monetary and financial instabilities management in the finances.

**Acknowledgement**

This research article is a result of our scientific collaborations with a number of leading researchers in the economics and finances in the Western Europe, North America, Asia and Australia, which have been conducted over the last *25* years. Authors would like to thank Dr. Ben Shalom Bernanke, *Chairman of the Board of Governors of the Federal Reserve System* for giving us his kind permission to obtain a copy of his *Ph. D. thesis* from the *Massachusetts Institute of Technology* in the *U.S.A.*; the insightful presentation titled: "What should economists and policymakers learn from the financial crisis?" at the *London School of Economics and Political Sciences* in London, *U.K.*; the thoughtful discussions on the policies towards the financial and monetary stabilities by the *Federal Reserve System* in the U.S.A.; and his tireless personal efforts toward the promotion of the economic and financial knowledge globally. We think that a set of our original ideas, which was presented and discussed in our research article, definitely requires more theoretical investigation in view of the complexity of considered research problems and the novelty of obtained research results.

*E-mail:     dimitri.ledenyov@my.jcu.edu.au .